\documentclass[twocolumn,pre,aps,showpacs,showkeys,amsmath]{revtex4}
\usepackage{graphicx}
\usepackage{adjustbox}
\usepackage{multirow}

\begin{document}

\title{Effect of Adult Neurogenesis on Sparsely Synchronized Rhythms of The Granule Cells in The Hippocampal Dentate Gyrus}
\author{Sang-Yoon Kim}
\email{sykim@icn.re.kr}
\author{Woochang Lim}
\email{wclim@icn.re.kr}
\affiliation{Institute for Computational Neuroscience and Department of Science Education, Daegu National University of Education, Daegu 42411, Korea}

\begin{abstract}
We are concerned about the main encoding granule cells (GCs) in the hippocampal dentate gyrus (DG). Young immature GCs (imGCs) appear through adult neurogenesis.
In comparison to the mature GCs (mGCs) (born during development), the imGCs show high activation due to lower firing threshold. On the other hand, they receive low excitatory drive from the entorhinal cortex via perforant paths and from the hilar mossy cells with lower connection probability $p_c~(=20~x~\%)$ ($x:$ synaptic connectivity fraction; $ 0 \leq x \leq 1$) than the mGCs with the connection probability $p_c~(=20~\%)$. Thus, the effect of low excitatory innervation (reducing activation degree) for the imGCs counteracts the effect of their high excitability. We consider a spiking neural network for the DG, incorporating both the mGCs and the imGCs. With decreasing $x$ from 1 to 0, we investigate the effect of young adult-born imGCs on the sparsely synchronized rhythms (SSRs) of the GCs (mGCs, imGC, and whole GCs). For each $x$, population and individual firing behaviors in the SSRs are characterized in terms of the amplitude measure ${\cal M}_a^{(X)}$ ($X=m,~im,~w$ for the mGCs, the imGCs, and the whole GCs, respectively) (representing the population synchronization degree) and the random phase-locking degree ${\cal L}_d^{(X)}$ (characterizing the regularity of individual single-cell discharges), respectively. We also note that, for $0 \leq x \leq 1,$ the mGCs and the imGCs exhibit pattern separation (i.e., a process of transforming similar input patterns into less similar output patterns) and pattern integration (making association between patterns), respectively. Quantitative relationship between SSRs and pattern separation and integration is also discussed.
\end{abstract}


\maketitle

\section{Introduction}
\label{sec:INT}
The hippocampus, consisting of the dentate gyrus (DG) and the subregions CA3 and CA1, plays important roles in memory formation, storage, and retrieval
(e.g., episodic and spatial memory) \cite{Gluck,Squire}. Here, we are concerned about the DG which is the gateway to the hippocampus. Its excitatory granule cells (GCs) receive excitatory inputs from the entorhinal cortex (EC) via the perforant paths (PPs). As a preprocessor for the CA3, the principal GCs perform pattern separation on the input patterns from the EC by sparsifying and orthogonalizing them, and send the pattern-separated outputs to the pyramidal cells in the CA3 via the mossy fibers (MFs)
\cite{Marr,Will,Mc,Rolls1,Rolls2a,Rolls2b,Treves1,Treves2,Treves3,Oreilly,Schmidt,Rolls3,Knier,Myers1,Myers2,Myers3,Scharfman,Yim,Chavlis,Kassab,PS1,PS2,PS3,PS4,PS5,PS6,PS7,PS}.
Then, the sparse, but strong MFs play a role of ``teaching inputs,'' causing synaptic plasticity between the pyramidal cells in the CA3.
Thus, a new pattern may be stored in modified synapses. In this way, pattern separation (transforming a set of input patterns into sparser and orthogonalized patterns) in the DG may facilitate pattern storage in the CA3.

The main encoding GCs in the DG are grouped into the lamellar clusters \cite{Cluster1,Cluster2,Cluster3,Cluster4}. In each cluster, both one inhibitory basket cell (BC) and one inhibitory HIPP (hilar perforant path-associated) cell exist, together with excitatory GCs. During pattern separation, the GCs exhibit sparse firing activity through the winner-take-all competition \cite{WTA1,WTA2,WTA3,WTA4,WTA5,WTA6,WTA7,WTA8,WTA9,WTA10,WTA}. Only strongly active GCs survive under the feedback inhibitory inputs from the BC and the HIPP cell. We note that, sparsity (arising from strong feedback inhibition) has been considered to improve the pattern separation efficacy \cite{Treves3,Oreilly,Schmidt,Rolls3,Knier,Myers1,Myers2,Myers3,Scharfman,Chavlis,Kassab,PS, WTA}.

Most distinctly, adult neurogenesis occurs in the DG, which leads to appearance of new young immature GCs (imGCs) during adulthood.
Pioneering studies of Altman in adult rat and cat brains for the adult neurogenesis were made decades ago in the 1960s \cite{NG1,NG2,NG3}.
Since then, adult neurogenesis has been found to be a robust phenomenon, occurring in most mammals, mainly in the subgranular zone of the DG and the subventricular zone of the lateral ventricles \cite{NG4,NG5,NG6}. The new imGCs born in the subgranular zone migrate into the granular layer of the DG.
Thus, the whole population of GCs consists of mature GCs (mGCs) born during the development and adult-born imGCs.
In comparison to the mGCs, the young adult-born imGCs are found to exhibit marked properties such as high excitability, weak inhibition, and low excitatory innervation \cite{NG7,NG8,NG9,NG10,NG11}.

\begin{figure*}
\includegraphics[width=1.5\columnwidth]{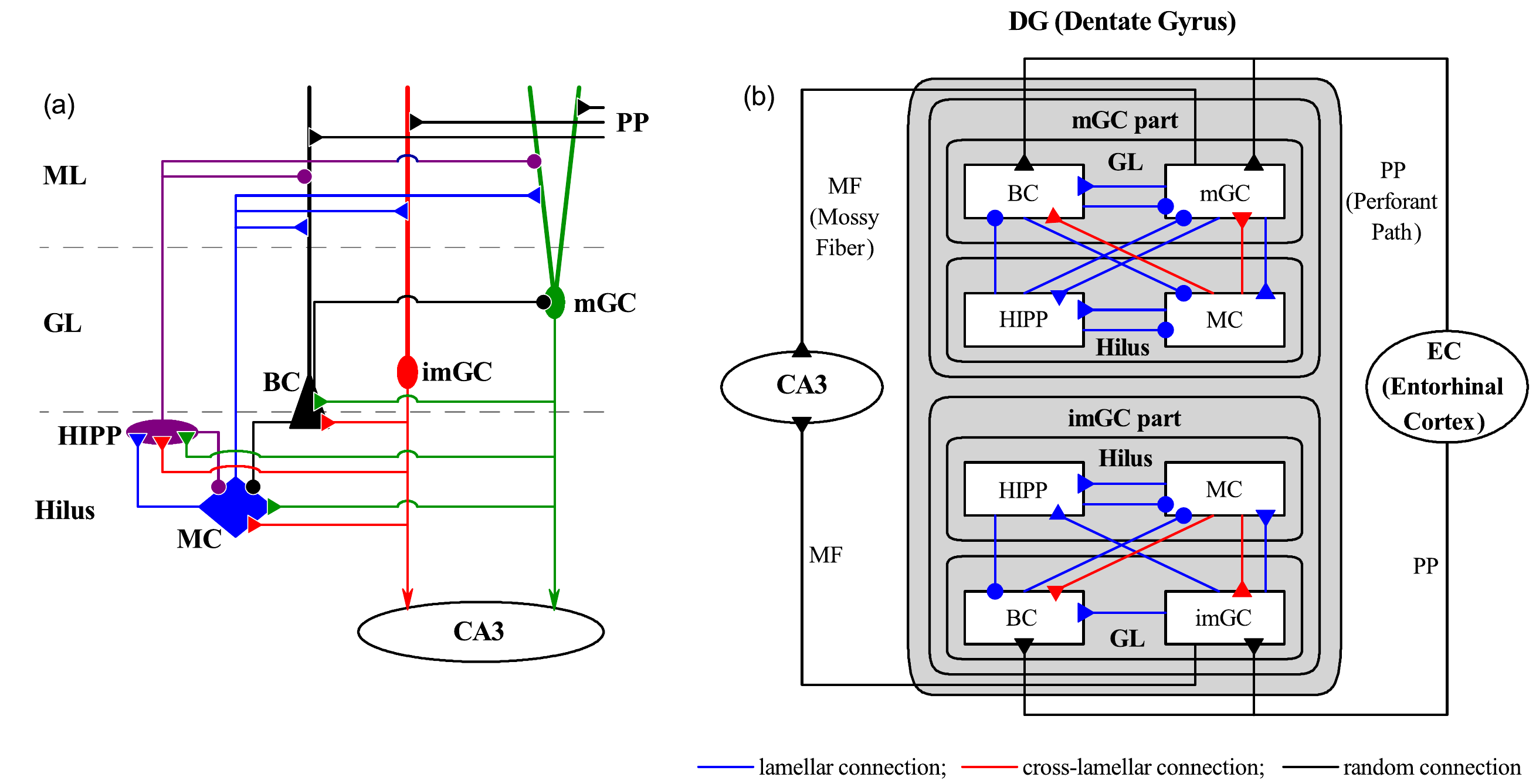}
\caption{Spiking neural network for the hippocampal dentate gyrus (DG). (a) Schematic representation of of major cells and synaptic connections in our DG network incorporating both mature GCs (mGCs) and adult-born immature GCs (imGCs). Fraction of the imGCs is 10 $\%$ in the whole population of GCs. Note that there are no inhibitory inputs into the imGCs, in contrast to the case of mGCs. Here, BC, MC, HIPP, PP, GL, and ML represent the basket cell, the mossy cell, the hilar perforant path-associated cell, perforant path, granular layer, and molecular layer, respectively. (b) Box diagram for our DG network with 3 types of synaptic connections. Blue, red, and black lines represent lamellar, cross-lamellar, and random connections, respectively.
}
\label{fig:DGN}
\end{figure*}

In this paper, we consider a spiking neural network for the adult neurogenesis in the DG, including both mGCs and imGCs (fraction of the imGCs is 10 $\%$)
where the effect of adult-born imGCs on pattern separation was studied \cite{NG-PS}.  In our DG network, both high excitability and low excitatory innervation
for the imGCs are considered; approximately no inhibition is provided to the imGCs. The imGCs exhibit high activation because of lower firing threshold, while they receive low excitatory drive from the entorhinal cortex (EC) via perforant paths (PPs) and from the hilar mossy cells (MCs) with lower connection probability $p_c~(=20~ x~\%)$ ($x:$ synaptic connectivity fraction; $ 0 \leq x \leq 1$) than the mGCs with the connection probability $p_c~(=20~\%)$. Thus, the effect of low excitatory innervation (decreasing activation degree) for the imGCs counteracts the effect of their high excitability.

We are concerned about population rhythms of the GCs in the DG. Sparsely synchronized rhythm (SSR) was found to appear in the presence of only mGCs
(without imGCs) during pattern separation via winner-take-all competition \cite{PS,SSR}. Here, with decreasing $x$ (synaptic connectivity fraction)
from 1 to 0, we investigate the effect of adult neurogenesis on SSRs of the GCs (mGCs, pmGC, and whole GCs) in our DG spiking neural network.
For each $x$, population and individual firing behaviors of the mGCs, the imGCs, and the whole GCs in their SSRs are characterized in terms of the amplitude measure ${\cal M}_a^{(X)}$ ($X=m,~im,~w$ for the mGCs, the imGCs, and the whole GCs, respectively) (denoting the population synchronization degree) and the random phase-locking degree ${\cal L}_d^{(X)}$ (characterizing the regularity of individual single-cell firings), respectively \cite{AM,SSR}. 

For $0 \leq x \leq 1,$ the mGCs and the whole GCs were found to exhibit pattern separation (i.e., a process of transforming similar input patterns into less similar output patterns), while the imGCs were found to show pattern integration (making association between patterns); efficacy of pattern separation and pattern integration was characterized in terms of pattern separation degree ${\cal S}_d^{(X)}$ ($X=m$ and $w$) and pattern integration degree ${\cal I}_d$ ($X=im$), respectively \cite{PS,NG-PS}. Here, quantitative relationship between ${\cal M}_a^{(X)}$ and ${\cal L}_d^{(X)}$ of the SSRs and ${\cal S}_d^{(X)}$ and ${\cal I}_d$ of pattern separation and integration is also discussed.

This paper is organized as follows. In Sec.~\ref{sec:DGN}, we describe a spiking neural network for the adult neurogenesis in the hippocampal DG. Then, in the main Sec.~\ref{sec:SSR}, we investigate the effect of the adult neurogenesis on SSRs of the GCs (mGCs, imGCs, whole GCs) by varying $x$ (synaptic connectivity fraction)
from 0 to 1. Also, we study  quantitative association between the SSRs and the efficacy of pattern separation and integration. Finally, we give summary and discussion in Sec.~\ref{sec:SUM}.

\section{Spiking Neural Network for The Adult Neurogenesis in The Dentate Gyrus}
\label{sec:DGN}
In this section, we describe our spiking neural network for the adult neurogenesis in the hippocampal DG.
Based on the anatomical and the physiological properties described in \cite{Myers1,Myers2,Chavlis}, we developed the DG spiking neural networks in the works for the winner-take-all competition \cite{WTA}, the SSR \cite{SSR}, and the pattern separation \cite{PS}. In our DG spiking neural network for the adult neurogenesis
\cite{NG-PS}, the young adult-born imGCs and the mGCs are incorporated, and more synaptic connections with a high degree of anatomical and physiological realism are also included \cite{BN1,BN2}.

\subsection{Framework of Our DG Spiking Neural Network for The Adult Neurogenesis}
\label{subsec:SNN}
We first describe framework of our DG spiking neural network for the adult neurogenesis.
The granular layer (GL) and the hilus constitute the DG. The GL consists of the excitatory mGCs and imGCs and the inhibitory BCs. The hilus is composed of
the excitatory MCs and the inhibitory HIPP cells, whose axons project to the upper molecular layer (ML).
Figure \ref{fig:DGN}(a) shows schematic representation of major cells and synaptic connections in our DG network for adult neurogenesis.
In our DG network, fraction of imGCs is 10 $\%$ in the whole population of GCs. High excitability and low excitatory innervation of the imGCs are considered, but there are no inhibitory inputs into the imGCs \cite{NG7,NG8,NG9,NG10,NG11}.

Based on the anatomical information given in \cite{Myers1,Myers2,Myers3,Scharfman,Chavlis}, we choose the numbers of GCs, BCs, MCs, and HIPP cells in the DG and the EC cells. As in our prior works \cite{WTA,SSR,PS}, we consider a scaled-down spiking neural network. The total number of excitatory GCs ($N_{\rm GC}$) is 2,000, corresponding to $\frac {1}{500}$ of the $10^6$ GCs found in rats \cite{ANA1}. Fraction of the imGCs in the whole population of the GCs is $10~\%$;
the number of the mGCs (imGCs) is 1800 (200). The whole GCs (i.e., mGCs and imGCs) are grouped into the $N_c~(=20)$ lamellar clusters \cite{Cluster1,Cluster2,Cluster3,Cluster4}. Then, in each GC cluster, there are $n_{\rm {GC}}^{(c)}~(=100)$ GCs (i.e., 90 mGC and 10 imGCs) and one inhibitory BC
\cite{Myers2,Myers3,Scharfman}. Next, we consider the hilus \cite{Hilus1,Hilus2,Hilus3,Hilus4,Hilus5,Hilus6,Hilus7}.
In our scaled-down DG network, we choose the number of MCs and the number of HIPP cells as $N_{\rm MC}=60$ and $N_{\rm HIPP}=20,$ respectively. Hence, in each cluster, there are $n_{\rm {MC}}^{(c)}~(=3)$ MCs and one HIPP cell \cite{Myers2,Myers3,Scharfman}.
Also, the number of EC cells (projecting the excitatory inputs to the mGCs, the imGCs, and the BCs through the PPs via random connections) in our scaled-down
neural network is $N_{\rm EC}=400$, and their activation degree is chosen as 10$\%$ \cite{ANA4}.
Each active EC cell is modeled in terms of the Poisson spike train with frequency of 40 Hz \cite{ANA5}.

Figure \ref{fig:DGN}(b) shows the box diagram for our DG network with 3 types of lamellar (blue), cross-lamellar (red), and random (black) synaptic connections
The EC provides the external excitatory inputs randomly to the mGCs, the imGCs, and the inhibitory BCs (with dendrites extending to the outer ML) through PPs \cite{Myers1,Myers2,Myers3,Scharfman,Chavlis}. Thus, both the mGCs and the imGCs receive direct excitatory EC input through PPs (EC $\rightarrow$ mGC and imGCs) via random connections. The connection probability $p_c$ for EC $\rightarrow$ mGC and BC is 20 $\%$, while $p_c$ for EC $\rightarrow$ imGC is decreased to $20~x$ $\%$ [$x$ (synaptic connectivity fraction); $ 0 \leq x \leq 1 $] because of low excitatory innervation. Furthermore, only the mGCs receive indirect feedforward inhibitory input, mediated by the BCs (EC $\rightarrow$ BC $\rightarrow$ mGC).

In the GL, the whole GCs (i.e., both the mGCs and the imGCs) are grouped into lamellar clusters \cite{Cluster1,Cluster2,Cluster3,Cluster4}, and one inhibitory BC exists in each cluster. Here, the BC (receiving excitation from the whole GCs in the same cluster) provides the feedback inhibition to all the mGCs in the same cluster via lamellar connections. In the hilus, we also consider lamellar organization for the MCs and HIPP cells \cite{Myers2,Myers3,Scharfman,Hilus6}. As in the case of BC, the HIPP cell receives excitation from the whole GCs in the same cluster, and projects the feedback inhibition to all the mGCs in the same cluster through lamellar connections.

In our DG network, the MCs play the role of ``controller'' for the activities of the two feedback loops of mGC-BC and mGC-HIPP. Each MC in a cluster receives excitation from all the GCs in the same cluster (lamellar connection), while it makes excitatory projection randomly to the mGCs and the imGCs in other clusters via cross-lamellar connections \cite{Hilus6}. The connection probability $p_c$ for MC $\rightarrow$ mGC is 20 $\%$, while $p_c$ for MC $\rightarrow$ imGC is decreased to $20~x$ $\%$ ($ 0 \leq x \leq 1 $) because of low excitatory innervation.

The MCs control the activities of the feedback loops of mGC-BC and mGC-HIPP. Each MC in a cluster receives inhibition from the BC and the HIPP cell in the same cluster (lamellar connection). Then, the MCs in the cluster project excitation to the BCs in other clusters through cross-lamellar connections (the connection probability $p_c$ for MC $\rightarrow$ BC is 20 $\%$) \cite{Hilus6}, while they provide excitation to the HIPP cell in the same cluster (lamellar connection).
Finally, the HIPP cell disinhibits the BC in the same cluster (lamellar connection for HIPP $\rightarrow$ BC); there are no reverse synaptic connections for HIPP $\rightarrow$ BC \cite{BN1,BN2}.

The mGCs in a cluster show sparse firing activity through the winner-take-all competition \cite{WTA1,WTA2,WTA3,WTA4,WTA5,WTA6,WTA7,WTA8,WTA9,WTA10,WTA}. Only strongly active mGCs can survive under the feedback inhibition from the BC and the HIPP cell in the same cluster. Here, the activities of the BC and the HIPP cell are controlled by the controller MCs; in the case of BC, the HIPP cell also disinhibits it. In contrast, the imGCs receive no inhibition. Particularly, because of  their low firing threshold, they become highly active, in contrast to the case of mGCs \cite{NG7,NG8,NG9,NG10}. However, when taking into consideration their low excitatory innervation from the EC cells and the MCs, their firing activity is reduced \cite{NG11}.

\subsection{Single Neuron Models and Synaptic Currents in Our DG Spiking Neural Network}
\label{subsec:SNSC}
As elements of our DG spiking neural network, we choose leaky integrate-and-fire (LIF) spiking neuron models with additional afterhyperpolarization (AHP) currents, determining refractory periods. This LIF spiking neuron model is one of the simplest spiking neuron models \cite{LIF}. Due to its simplicity, it can be easily analyzed and simulated.

Our DG network consists of 5 populations of mGCs, imGCs, BCs, MCs, and HIPP cells. The state of a neuron in each population is characterized by its membrane potential. Then, time-evolution of the membrane potential is governed by 4 types of currents into the neuron; the leakage current, the AHP current, the external constant current, and the synaptic current.

We note that the equation for a single LIF neuron model (without the AHP current and the synaptic current) describes a simple parallel resistor-capacitor (RC) circuit. Here, the 1st type of leakage current is because of the resistor and the integration of the external current is because of the capacitor which is in parallel to the resistor. When its membrane potential reaches a threshold, a neuron fires a spike, and then the 2nd type of AHP current follows. As the decay time of the AHP current is increased, the refractory period becomes longer. Here, we consider a subthreshold case where the 3rd type of external constant current is zero \cite{Chavlis}.

Detailed explanations on the leakage current and the AHP current, associated with each type of single neuron (mGC, imGC, BC, MC, and HIPP cell), are given in Appendix \ref{app:A}. The parameter values of the capacitance $C_X$, the leakage current $I_L^{(X)}(t)$, and the AHP current $I_{AHP}^{(X)}(t)$ are the same as those
in our prior DG networks \cite{WTA,SSR,PS}, and refer to Table I in \cite{WTA}. These parameter values are based on physiological properties of the GC, BC, MC,
and HIPP cell \cite{Chavlis,Hilus3}.

We note that, the GC in Table 1 in \cite{WTA} corresponds to the mGC. The imGCs also have the same parameter values as those of the mGC, except
for the leakage reversal potential $V_L$. The mGC with $V_L=-75$ mV exhibits a spiking transition when passing a threshold $I^*=80$ pA. Here, we consider a case that the imGC has an increased leakage reversal potential of $V_L=-72$ mV, which could lead to intrinsic high excitability. Then, it shows a firing transition when passing $I^*=69.7$ pA. In this way, the imGC may have a lower firing threshold \cite{NG7,NG8,NG9,NG10}, which is well shown
in Fig.~2 for the $f-I$ (i.e., firing rate-current) curves of the mGC (red curve) and the imGC (blue curve) in \cite{NG-PS}.

Next, we consider the 4th type of synaptic current. Detailed explanations on the synaptic current are given in Appendix B; here, we give a brief and clear explanation on it. There are 3 kinds of synaptic currents from a presynaptic source population to a postsynaptic neuron in the target population; 2 kinds of excitatory AMPA and NMDA receptor-mediated synaptic currents and one type of inhibitory GABA receptor-mediated synaptic current.
In each $R$ (AMPA, NMDA, and GABA) receptor-mediated synaptic current, the synaptic conductance is given by the product of the synaptic strength per synapse, the average number of afferent synapses (connected to a postsynaptic neuron), and fraction of open ion channels.

The postsynaptic ion channels are opened via binding of neurotransmitters (emitted from the source population) to receptors in the target
population. The time course of fraction of open ion channels is given by a sum of ``double-exponential'' functions over presynaptic spikes.
The double-exponential function, corresponding to contribution of a presynaptic spike, is controlled by the synaptic rising time constant, the synaptic decay time
constant, and the synaptic latency time constant; for details, refer to Eq.~(\ref{eq:ISyn4}) in Appendix B.

The parameter values for the synaptic strength per synapse, the synaptic rising time constant, the synaptic decay time constant, the synaptic latency time constant, and the synaptic reversal potential for the synaptic currents into the GCs, the BCs, the MCs, and the HIPP cells are given
in Tables I-III in \cite{NG-PS}. These parameter values are also based on the physiological properties of the relevant cells \cite{Chavlis,SynParm1,SynParm2,SynParm3,SynParm4,SynParm5,SynParm6,SynParm7,SynParm8}.

All of our source codes for computational works were written in C programming language. Numerical integration of the governing equation for the time-evolution of states of individual spiking neurons is done by employing the 2nd-order Runge-Kutta method with the time step 0.1 msec.

\begin{figure}
\includegraphics[width=1.0\columnwidth]{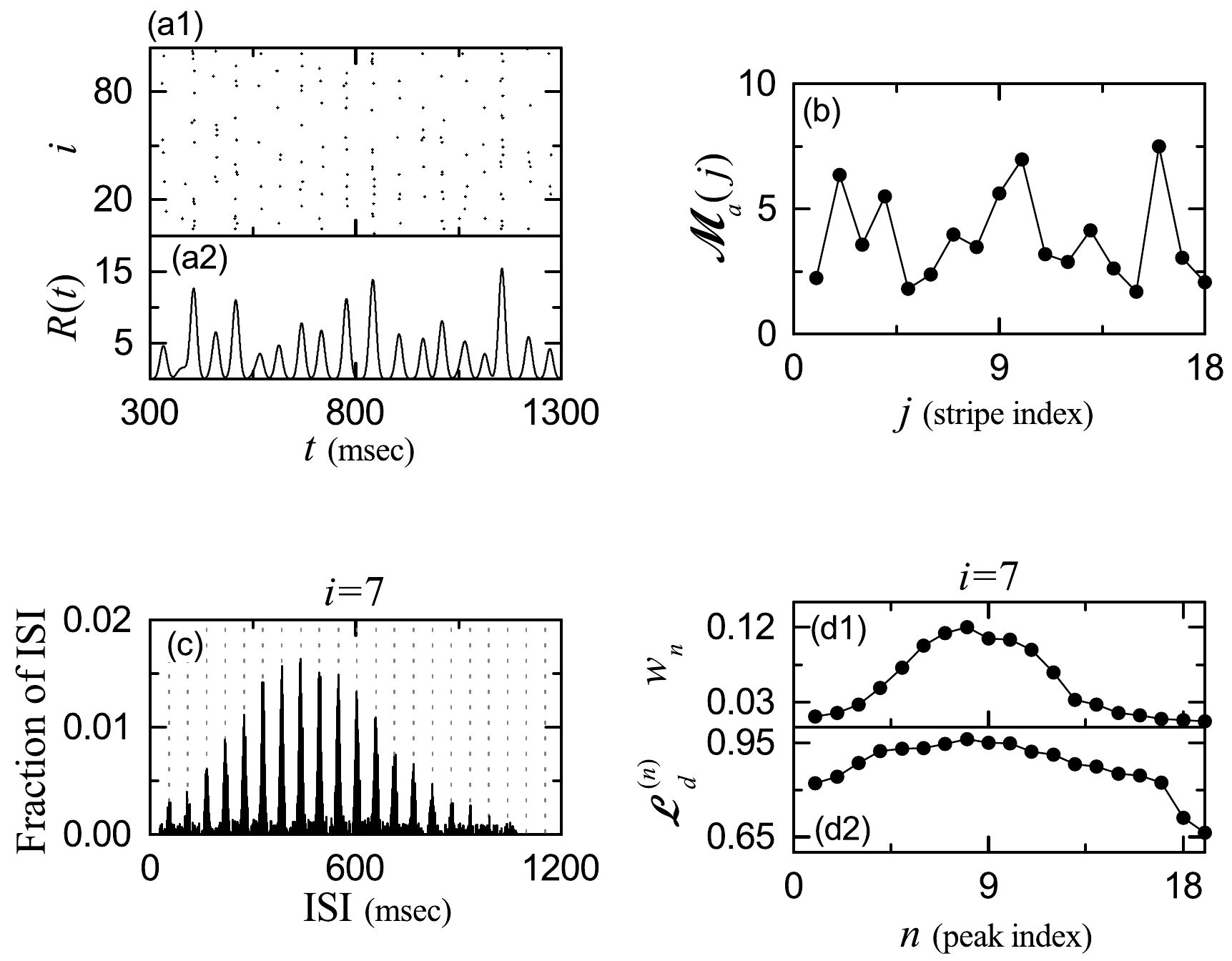}
\caption{Characterization of SSR in the presence of only the mGCs without imGCs. (a1) Raster plots of spikes of 120 active mGCs. (a2)
instantaneous population spike rate (IPSR) $R(t)$ of active mGCs. Band width for $R(t)$: $h=20$ msec. (b) Plot of amplitude measure ${\cal M}_a(j)$ of the IPSR $R(t)$ versus $j$ (spiking stripe). (c) ISI histogram of the 7th active mGC; bin size = 2 msec. Vertical dotted lines in (c) represent the integer multiples of the global period $T_G$ (= 54.9 msec) of $R(t)$. Plots of (d1) normalized weight $w_n$ and (d2) random phase-locking degree ${\cal L}_d^{(n)}$ for the $n$th peak of the ISI histogram for the 7th active mGC versus $n$ (peak index).
}
\label{fig:SSR}
\end{figure}

\section{Effect of Adult Neurogenesis on Sparsely Synchronized Rhythms}
\label{sec:SSR}
In this section, we study the effect of adult neurogenesis on the SSRs of the GCs (mGCs, imGCs, whole GCs) in our DG spiking neural network.
With decreasing $x$ (synaptic connectivity fraction) from 1 to 0, population and individual firing behaviors of the mGCs, the imGCs, and the whole GCs in their SSRs are investigated by employing the amplitude measure ${\cal M}_a^{(X)}$ ($X=m,~im,~w$ for the mGCs, the imGCs, and the whole GCs, respectively) (denoting the population synchronization degree) \cite{AM} and the random phase-locking degree ${\cal L}_d^{(X)}$ (characterizing the regularity of individual single-cell firings) \cite{SSR,PS}, respectively. For $0 \leq x \leq 1$ the mGCs and the imGCs were found to exhibit pattern separation and pattern integration, respectively \cite{NG-PS}. We also discuss quantitative relationship between the SSRs and the pattern separation and integration.

\subsection{Characterization of Sparsely Synchronized Rhythm in The Presence of Only The mGCs without The imGCs}
\label{subsec:mGC}
We first consider the homogeneous population of mGCs (without the imGCs) \cite{SSR}.
Population firing activity of the active mGCs may be well visualized in the raster plot of spikes which is a collection of spike trains of individual active mGCs. Figure \ref{fig:SSR}(a1) shows the raster plot of spikes for 120 active mGCs (activation degree $D_a$ of the mGCs is 6 $\%$); for convenience, only a part from $t=300$ to 1,300 msec is shown in the raster plot of spikes. We note that sparsely synchronized stripes (composed of sparse spikes and indicating population sparse synchronization) appear successively.

As a population quantity showing collective behaviors, we employ an IPSR (instantaneous population spike rate) which may be obtained from the raster plot of spikes
\cite{W_Review,Sparse1,Sparse2,Sparse3,FSS,SM}. To get the smooth IPSR, we employ the kernel density estimation (kernel smoother) \cite{Kernel}. Each spike in the raster plot is convoluted (or blurred) with a kernel function $K_h(t)$ to get a smooth estimate of IPSR $R(t)$:
\begin{equation}
R(t) = \frac{1}{N_a} \sum_{i=1}^{N_a} \sum_{s=1}^{n_i} K_h (t-t_{s,i}),
\label{eq:IPSR}
\end{equation}
where $N_a$ is the number of the active mGCs, $t_{s,i}$ is the $s$th spiking time of the $i$th active mGC,
$n_i$ is the total number of spikes for the $i$th active mGC, and we use a Gaussian kernel function of band width $h$:
\begin{equation}
K_h (t) = \frac{1}{\sqrt{2\pi}h} e^{-t^2 / 2h^2}, ~~~~ -\infty < t < \infty,
\label{eq:Gaussian}
\end{equation}
where the band width $h$ of $K_h(t)$ is 20 msec. The IPSR $R(t)$ is also shown in Fig.~\ref{fig:SSR}(a2).
We note that the IPSR $R(t)$ exhibits synchronous oscillation with the population frequency $f_p~(= 18.2$ Hz).
The population frequency $f_p$ is given by the reciprocal of the global period $T_G$ (i.e., $f_p = 1 / T_G$)
which corresponds to the average ``intermax'' interval (i.e., average interval between neighboring maxima) in the IPSR $R(t)$.
Here, we get $N_{\rm IMI}$ (= 545) intermax intervals during the stimulus period $T_s$ ($ = 3 \cdot 10^4$ msec), and get
their average value (i.e., global period) $T_G$ (= 54.9 msec).
In this way, SSR with $f_p~(= 18.2$ Hz) appears in the (homogeneous) population of active mGCs.

The amplitude of the IPSR $R(t)$ may represent synchronization degree of the SSR. Here, we characterize the synchronization degree of the SSR in terms of the amplitude measure ${\cal M}_a$, given by the time-averaged amplitude of $R(t)$ \cite{AM}:
\begin{equation}
  {\cal M}_a = {\overline {{\cal M}_a(j)}}; {\cal M}_a(j) = \frac {[R_{\rm max}^{(j)}(t) - R_{\rm min}^{(j)}(t)]} {2},
\label{eq:AM}
\end{equation}
where the overline represents time average, ${\cal M}_a(j)$ is the amplitude measure in the $j$th global cycle (corresponding to the $j$th spiking stripe), and $R_{\rm max}^{(j)}(t)$ and $R_{\rm min}^{(j)}(t)$ are the maximum and the minimum of $R(t)$ in the $j$th global cycle, respectively. As ${\cal M}_a$ increases (i.e., the time-averaged amplitude of $R(t)$ is increased), synchronization degree of the SSR becomes higher.
Figure \ref{fig:SSR}(b) shows plot of the amplitude ${\cal M}_a(j)$ versus the spiking stripe index $j$ (corresponding to the global cycle index). We follow the
546 stripes during the stimulus period $T_s$ ($ = 3 \cdot 10^4$ msec), and the amplitude measure ${\cal M}_a$ [corresponding to the time-averaged amplitude $\overline { {\cal M}_a(j) }$] is thus found to be 3.83.

Next, we consider the individual firing behavior of the active mGCs. For each active mGC, we get the inter-spike-interval (ISI) histogram by collecting the ISIs during the stimulus period $T_s$ ($= 3 \cdot 10^4$ msec). Each active mGC exhibits intermittent spikings, phase-locked to $R(t)$ at random multiples of its global period $T_G$ (= 54.9 msec). This is in contrast to the case of full synchronization where only one dominant peak appears at the global period $T_G$; all cells fire regularly at each global cycle without skipping. As a result of random spike skipping, there appear 19 distinct multiple peaks at the integer multiples of $T_G$ in the ISI histogram. These peaks are called as the random-spike-skipping peaks. Then, we get the population-averaged ISI histogram by averaging the individual ISI histograms for all the active mGCs. In this case, the population-averaged ISI ($\langle {\rm ISI} \rangle$) of all the active mGCs in the population-averaged ISI histogram is 471.7 msec. Then, the population-averaged mean firing rate (MFR) $\langle f_i \rangle$, given by the reciprocal of  $\langle {\rm ISI} \rangle$
(i.e., $\langle f_i \rangle = 1/ \langle {\rm ISI} \rangle$), is 2.12 Hz, which is much less than the population frequency $f_p~(=18.2$ Hz) of the SSR, in contrast to the case of full synchronization where the population-averaged MFR is the same as the population frequency.

As an example, we consider the case of the 7th ($i=7$) active mGC. Its ISI histogram is shown in Fig.~\ref{fig:SSR}(c). In this case, the 8th-order peak is the highest one, and hence spiking may occur most probably after 7-times spike skipping.
The $n$th-order random-spike-skipping peak in the ISI histogram is located as follows:
\begin{eqnarray}
 && (n-\frac {1}{2})~ T_G < {\rm ISI} < (n+ \frac {1}{2})~ T_G~~~~~{\rm for}~n \geq 2, \\
 && 0 < {\rm ISI} < {\frac {3}{2}}~ T_G~~~~~{\rm for}~n=1.
\label{eq:Peak}
\end{eqnarray}
For each $n$th-order peak, we obtain the normalized weight $w_n$, given by:
\begin{equation}
w_n = \frac {N_{\rm ISI}^{(n)}} {N_{\rm ISI}^{(tot)}},
\end{equation}
where $N_{\rm ISI}^{(tot)}$ is the total number of ISIs obtained during the stimulus period ($T_s$ $= 3 \cdot 10^4$ msec) and $N_{\rm ISI}^{(n)}$ is the number of the ISIs in the $n$th-order peak. Figure \ref{fig:SSR}(d1) shows plot of $w_n$ versus $n$ (peak index) for all the 19 peaks in the ISI histogram of the 7th ($i=7$) active mGC. For example, the highest 8th-order peak has $w_8=0.12$.

We now consider the sequence of the ISIs, $\{ {\rm ISI}_j^{(n)},~ j=1,\dots,N_{\rm ISI}^{(n)} \}$, within the $n$th-order peak, and get the random
phase-locking degree ${\cal L}_d^{(n)}$ of the $n$th-order peak (representing how well intermittent spikes make phase-locking to the IPSR $R(t)$
at $t = n T_G$). As in the case of the pacing degree \cite{SM}, we give a phase $\psi$ to each ${\rm ISI}_j^{(n)}$ via linear interpolation:
\begin{equation}
  \psi(\Delta {\rm ISI}_j^{(n)}) = \frac {\pi} {T_G}~\Delta {\rm ISI}_j^{(n)}~~~~~{\rm for}~n \geq 2,
\label{eq:phase1}
\end{equation}
where $\Delta {\rm ISI}_j^{(n)} = {\rm ISI}_j^{(n)} - n ~T_G$, leading to $ - {\frac {T_G}{2}}  < \Delta {\rm ISI}_j^{(n)} < {\frac {T_G}{2}} $. However, for $n=1$, $\psi$ changes depending on whether the ISI lies in the left or the right part of the 1st-order peak:
\begin{equation}
\psi(\Delta {\rm ISI}_j^{(1)}) = \left\{
\begin{array}{l}
\frac {\pi} {2~T_G}~\Delta {\rm ISI}_j^{(1)}~~~~~{\rm for}~ - {T_G} < \Delta {\rm ISI}_j^{(1)} < 0, \\
\frac {\pi} {T_G}~\Delta {\rm ISI}_j^{(1)}~~~~~{\rm for}~ 0 < \Delta {\rm ISI}_j^{(1)} < {\frac {T_G}{2}},
\end{array}
\right.
\label{eq:phase2}
\end{equation}
where $\Delta {\rm ISI}_j^{(1)} = {\rm ISI}_j^{(1)} - T_G.$

Then, the contribution of the ${\rm ISI}_j^{(n)}$ to the locking degree ${\cal L}_d^{(n)}$ is given by $\cos ( \psi_j^{(n)})$;
$\psi_j^{(n)} = \psi (\Delta {\rm ISI}_j^{(n)})$. An ${\rm ISI}_j^{(n)}$ makes the most constructive contribution to ${\cal L}_d^{(n)}$ for $\psi_j^{(n)}=0$, while it makes no contribution to ${\cal L}_d^{(n)}$ for $\psi_j^{(n)} = {\frac {\pi} {2}}$ or $-{\frac {\pi} {2}}$.
By averaging the matching contributions of all the ISIs in the $n$th-order peak, we get:
\begin{equation}
  {\cal L}_d^{(n)} = { \frac {1}{N_{\rm ISI}^{(n)}} } \sum_j^{N_{\rm ISI}^{(n)}}  \cos ( \psi_j^{(n)}).
\label{eq:LDn}
\end{equation}
Figure \ref{fig:SSR}(d2) shows plot of ${\cal L}_d^{(n)}$ versus $n$ (peak index) for the 19 random-spike-skipping peaks in the ISI histogram of the 7th active mGC.
For example, the highest 8th-order ($n=8$) peak has the maximum value of ${\cal L}_d^{(n)}$ (= 0.961).
Through weighted average of the random phase-locking degrees ${\cal L}_d^{(n)}$ of all the peaks, we obtain the (overall) random phase-locking degree ${\cal L}_d$
\begin{equation}
  {\cal L}_d = \sum_{n=1}^{N_p} w_n \cdot {\cal L}_d^{(n)}
             = {\frac {1} {N_{\rm ISI}^{\rm (tot)}} } \sum_{n=1}^{N_p} \sum_{j=1}^{N_{\rm ISI}^{\rm (tot)} } \cos ( \psi_j^{(n)}),
\label{eq:LD}
\end{equation}
where $N_p$ is the number of peaks in the ISI histogram.
We note that, ${\cal L}_d$ corresponds to the average of contributions of all the ISIs in the ISI histogram.
In the case of the 7th active mGC, the random phase-locking degree ${\cal L}_d$, characterizing the sharpness of all the peaks, is 0.92.
Hence, the mGCs make intermittent spikes which are well phase-locked to $R(t)$ at random multiples of its global period $T_G$.

We repeat the above process in the ISI histogram of each $i$th ($i=1,\dots,120$) active mGC and get its random phase-locking degree
${\cal L}_d(i)$. The range of $\{ {\cal L}_d(i) \} $ is [0.65, 1.23]. Then, the random phase-locking degree ${\cal L}_d$ of all the active mGCs ia given
by the average value (= 0.92) of the distribution $\{ {\cal L}_d(i) \}$.

\begin{figure*}
\includegraphics[width=1.5\columnwidth]{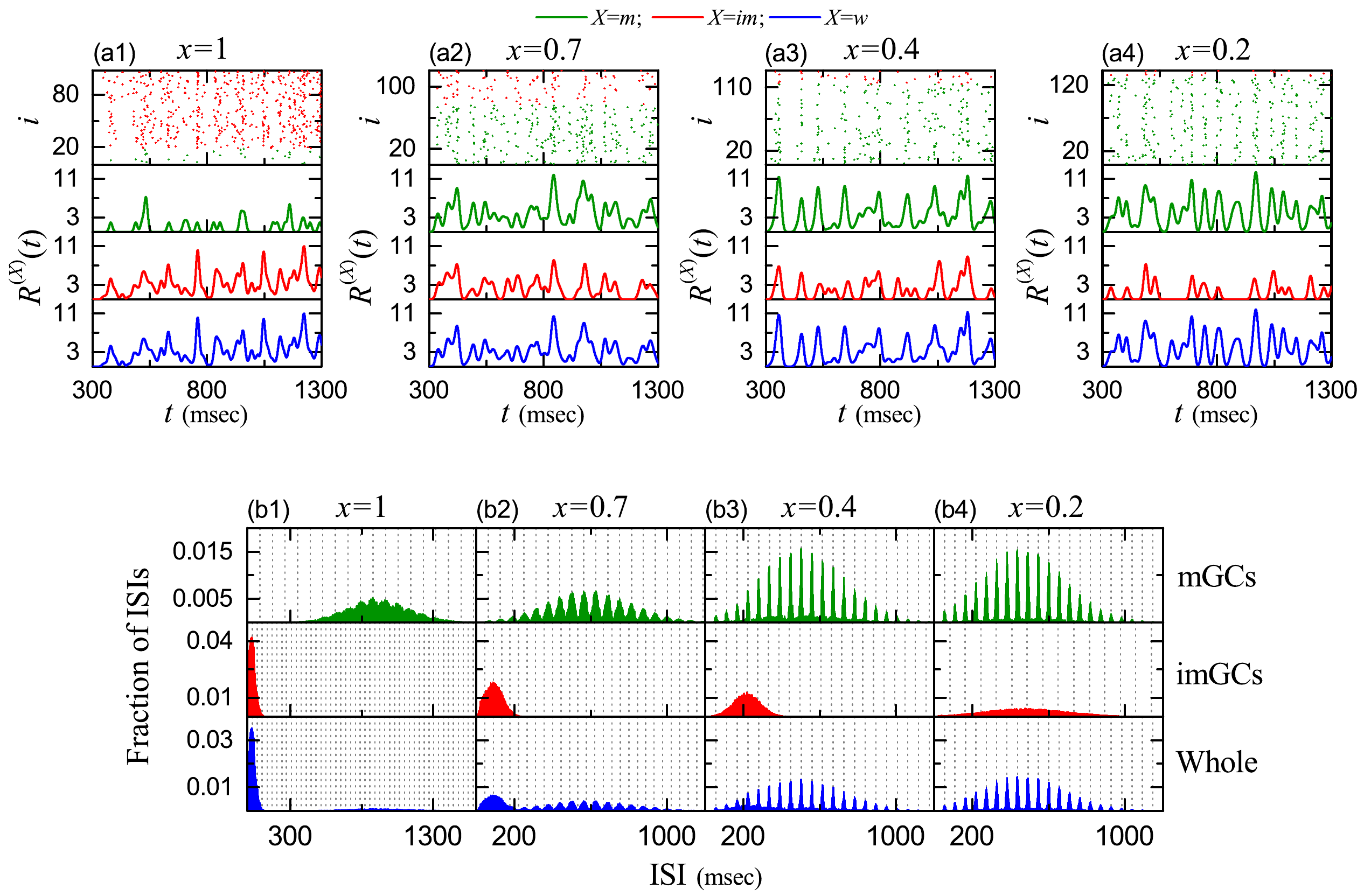}
\caption{SSR and multi-peaked ISI histogram in each case of the mGCs, the imGCs, and the whole GCs.
(a1)-(a4) Raster plots of spikes and IPSRs $R^{(X)}(t)$ for the active mGCs ($X=m$), the imGCs ($X=im$), and the whole GCs ($X=w$)
when $x$ (synaptic connectivity fraction) is 1.0, 0.7, 0.4, and 0.2, respectively.
(b1)-(b4) Population-averaged ISI histograms for the active mGC, imGC, and whole GCs when $x =$ 1.0, 0.7, 0.4, and 0.2, respectively;
bin size = 2 msec. Vertical dotted lines in (b1)-(b4) represent the integer multiples of the global period $T_G^{(X)}$ of $R^{(X)}(t)$.
In (a1)-(a4) and (b1)-(b4), mGCs, imGCs, and whole GCs are denoted in green, red, and blue color, respectively.
}
\label{fig:SSR1}
\end{figure*}

\subsection{Effect of The Adult-Born imGCs on Sparsely Synchronized Rhythms}
\label{subsec:imGC}
In this subsection, we consider a heterogeneous population, composed of mGCs and imGCs; fraction of the imGCs in the whole population is 10 $\%$.
As shown in Fig.~2 in \cite{NG-PS}, as a result of increased leakage reversal potential $V_L$, the imGC has lower firing threshold than the mGC (i.e., high excitability), which results in high activation of the imGC \cite{NG7,NG8,NG9,NG10}. We also note that, the imGC has low excitatory innervation from the EC cells and the hilar MCs, counteracting its high excitability \cite{NG11}. In the case of the mGC, the connection probability $p_c$ from the EC cells and the MCs to the mGC is 20 $\%$, while in the case of the imGC, $p_c$ is decreased to $20~x~\%$ [$x:$ synaptic connectivity fraction; $0 \leq x \leq 1$]. Due to low excitatory drive from the EC cells and the MCs, the activation degree of the imGC becomes reduced. With decreasing $x$ from 1 to 0, we investigate the effect of high excitability and low excitatory innervation for the imGC on the population and individual firing behaviors of the mGCs, the imGCs, and the whole GCs in their SSRs. We also note that, for $0 \leq x \leq 1$ the mGCs and the imGCs were found to exhibit pattern separation and pattern integration, respectively \cite{NG-PS}. Hence, we also study quantitative relationship between  SSRs and pattern separation and integration.

Here, as in the case of Fig.~\ref{fig:SSR}, we consider a long-term stimulus stage (300-30,300 msec) (i.e., the stimulus period $T_s=30,000$ msec), because long-term stimulus is necessary for analysis of dynamical behaviors. Population firing activity of the active mGCs and imGCs may be well visualized in the raster plot of spikes which is a collection of spike trains of individual active GCs. Figures \ref{fig:SSR1}(a1)-\ref{fig:SSR1}(a4) show the raster plots of spikes for the active mGCs (green) and imGCs (red) for $x=$ 1.0, 0.7, 0.4, and 0.2, respectively. For convenience, only a part from $t=300$ to 1,300 msec is shown in each raster plot of spikes. We note that sparsely synchronized stripes (composed of sparse spikes and indicating population sparse synchronization) appear successively; overall, the pacing degree between spikes in the spiking stripes is low. In the case of mGCs, with decreasing $x$ from 1 their spiking stripes become clearer, while in the case of imGCs their stripes become more smeared.

The instantaneous population spike rate [IPSR (showing population firing behavior)] may be obtained from the raster plot of spikes
[see Eq.~(\ref{eq:IPSR})]. The IPSRs $R^{(X)}(t)$ of the mGCs ($X=m:$ green), the imGCs ($X=im:$ red), and the whole GCs ($X=w:$ blue) are shown in Figs.~\ref{fig:SSR1}(a1)-\ref{fig:SSR1}(a4) for $x=$ 1.0, 0.7, 0.4 and 0.2, respectively. We note that $R^{(X)}(t)$ exhibit synchronous oscillations. But,
the average amplitude of $R^{(X)}(t)$ in each case of mGCs, imGCs, and whole GCs is smaller than that in the case of homogeneous population of only mGCs in Fig.~\ref{fig:SSR}(a2), and variations in the amplitudes are also large.

For $x=1$, imGCs fire spikings much more actively than mGCs because the imGCs have high excitability. On the other hand, firing activity of mGCs becomes much decreased due to strongly increased feedback inhibition from the BCs and the HIPP cells. Hence, in the case of $x=1$ the amplitude of $R^{(im)}(t)$ (red) of the imGCs is larger than that of $R^{(m)}(t)$ (green) of the mGCs. However, as $x$ is decreased from 1, firing activity of the imGCs becomes rapidly reduced (i.e., the effect of imGCs decreases rapidly) because of low excitatory innervation from the EC cells and the MCs. On the other hand, firing activity of mGCs becomes enhanced due to decrease in the feedback inhibition into the mGCs from the BCs and the HIPP cells. Thus, with decreasing $x$ from 1, the amplitude of $R^{(m)}(t)$ of the mGCs makes an increase because the pacing degree between spikes in each spiking stripe in the rater plot of spikes becomes better (i.e., the spiking stripes in the raster plot of spikes become clearer). In contrast, in the case of imGCs, the amplitude of $R^{(im)}(t)$ decreases because the pacing degree of spikes in the raster plot becomes worse (i.e., the spiking stripes in the raster plot of spikes become smeared). Thus, for example, for $x=0.2$ the amplitude of $R^{(m)}(t)$ becomes much larger than that of $R^{(im)}(t)$.

In the case of whole GCs, $R^{(w)}(t)$ (blue) gets an ``average'' value of $R^{(m)}(t)$ and $R^{(im)}(t)$;
$R^{(w)}(t) = [N_a^{(m)} / N_a^{(w)}] R^{(m)}(t) + [N_a^{(im)} / N_a^{(w)}] R^{(im)}(t)$ ($N_a^{(X)}$ is the number of active GCs in the $X$ population).
Thus, for any $x$, $R^{(w)}(t)$ follows the tendency of the larger one between $R^{(m)}(t)$ and $R^{(im)}(t)$; for example, for $x=1$ $R^{(w)}(t)$ is close to $R^{(im)}(t)$, while for other values of $x=$ 0.7, 0.4, and 0.2, it is near to $R^{(m)}(t).$

In addition to the (above) population firing activity, we also study the individual spiking activity of the active GCs. In each case of the mGCs ($X=m$), the imGCs ($X=im$), and the whole GCs ($X=w$), we get the ISI histogram for each active GC by collecting the ISIs during the stimulus period $T_s$ ($= 3 \cdot 10^4$ msec),
and then obtain the population-averaged ISI histogram by averaging the individual ISI histograms for all the active GCs.
Figures \ref{fig:SSR1}(b1)-\ref{fig:SSR1}(b4) show the population-averaged ISI histograms for $x=$ 1, 0.7, 0.4, and 0.2, respectively.

We first consider the case of $x=1$ in Fig.~\ref{fig:SSR1}(b1). For the mGCs (green), each active mGC exhibits intermittent spikings, phase-locked to $R^{(m)}(t)$ at random multiples of its global period $T_G^{(m)}$ (= 87.8 msec) [corresponding to the average ``intermax'' interval between neighboring maxima in $R^{(m)}(t)$]; vertical dotted lines represent integer multiples of the global period $T_G^{(m)}$ of $R^{(m)}(t)$. As a result of random spike skipping, there appear 12 multiple peaks in the ISI histogram. The middle 10th-order peak is the highest one, and hence spiking may occur most probably after 9-times spike skipping. This is in contrast to the case of full synchronization where only one dominant peak appears at the global period $T_G$ of the IPSR $R(t)$; all cells fire regularly at each global cycle without skipping. Next, we consider the case of imGCs (red). Its ISI histogram has a single peak near the global period $T_G^{(im)}~(=34.3$ msec) of the IPSR $R^{(im)}(t)$ and its distribution is broadly extended to $\sim 3~T_G^{(im)}$. The imGCs exhibit spikes mainly at $T_G^{(im)}$ (i.e., they fire mainly in each stripe), but they also show intermittent spikings at 2 $T_G^{(im)}$ or 3 $T_G^{(im)}$ (i.e., spike skippings also occur).

Overall, we consider the case of whole GCs (blue). In the case of $x=1$, spikes of the imGCs are dominant, as shown in Fig.~\ref{fig:SSR1}(a1). Thus, dominant major peak, associated with the imGCs, appears near the global period $T_G^{(w)}$ (= 34.3 msec) of the IPSR $R^{(w)}(t)$, while fractions of multiple peaks, related to the mGCs, are very small (not clearly seen).

As $x$ is decreased from 1 (i.e., considering low excitatory innervation for the imGCs), the effect of the imGCs becomes weaker.
In this case, the imGCs show more irregular spiking behaviors. Hence, their single-peaked ISI histograms become broader, as shown in
Figs.~\ref{fig:SSR1}(b2)-\ref{fig:SSR1}(b4) for $x=$ 0.7, 0.4, and 0.2, respectively. The order $n$ of peak also increases with decreasing $x$
($n=$ 2, 4, and 6 for $x=$ 0.7, 0.4, and 0.2, respectively). Hence, more spike skippings occur.
On the other hand, with decreasing $x$ from 1, the mGCs exhibit more regular spiking behaviors. Hence, their ISI histograms become clearer because multiple peaks become sharper and their heights become increased [see Figs.~\ref{fig:SSR1}(b2)-\ref{fig:SSR1}(b4)].

In the case of whole GCs, both the peak (associated with the imGCs) and the multiple peaks (related to the mGCs)
coexist in the histogram. However, their fractions vary depending on $x$. For $x=0.7,$ the effect of the imGCs becomes reduced, and hence
the height of the peak at $2~T_G^{(w)}$, associated with the imGCs, becomes decreased. In this case, multiple peaks, related to the mGCs, become clearly visible, because their fractions are increased. However, with further decrease in $x$, spikes of mGCs become more and more dominant. Thus, only the multiple peaks, related to the mGCs, are visible because fraction of the peak, associated with the imGCs, becomes very small.

\begin{figure}
\includegraphics[width=\columnwidth]{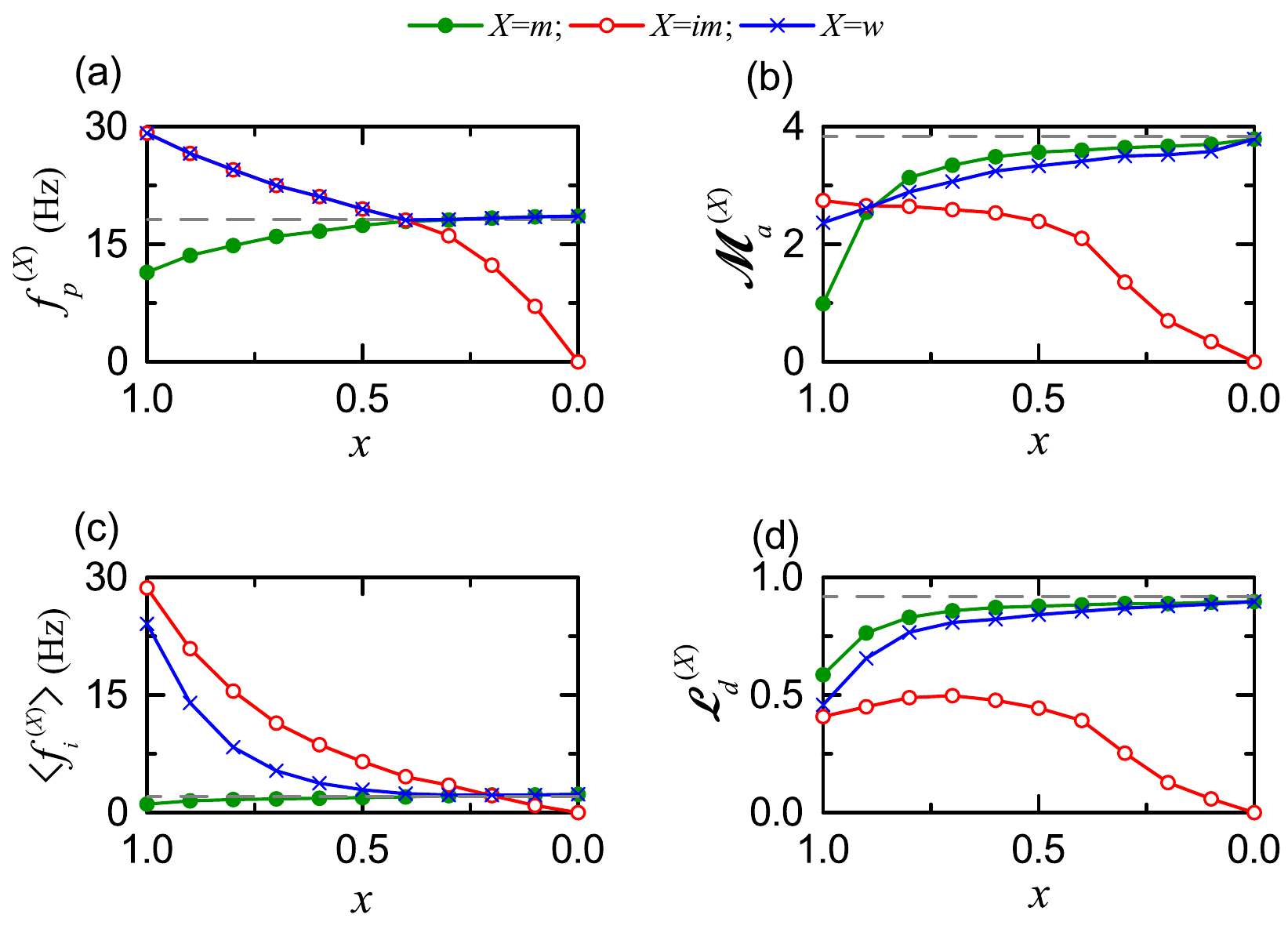}
\caption{Population and individual firing behaviors in the SSRs of the mGCs ($X=m$), the imGCs ($X=im$), and the whole GCs ($X=w$).
(a) Plots of the population frequencies $f_p^{(X)}$ versus $x$ (synaptic connectivity fraction). (b) Plots of the amplitude measures ${\cal M}_a^{(X)}$ versus $x$.
(c) Plots of the population-averaged mean firing rates $\langle f_i^{(X)} \rangle$ versus $x$.
(d) Plots of the  random-phase-locking degrees ${\cal L}_d^{(X)}$ versus $x$.
Horizontal dashed lines in (a)-(d) represent $f_p$ (= 18.2 Hz), $M_a$ (= 3.83), $\langle f_i \rangle$ (= 2.12 Hz), and ${\cal L}_d$ (= 0.92) in the presence of
only mGCs (without imGCs), respectively.
}
\label{fig:SSR2}
\end{figure}

From now on, in Fig.~\ref{fig:SSR2}, we quantitatively characterize population and individual firing behaviors in the SSRs of the mGCs ($X=m$), the imGCs ($X=im$), and the whole GCs ($X=w$). We first consider the population firing behaviors which are well shown in $R^{(X)}(t)$.
Figure \ref{fig:SSR2}(a) shows the plots of the population frequency $f_p^{(X)}$ [i.e., the average oscillating frequency of $R^{(X)}(t)$, corresponding
to the reciprocal of the global period $T_G^{(X)}$ of the SSRs for $X=m$ (green solid circles), $im$ (red open circles), and $w$ (blue crosses)]. For $x=1$, $f_p^{(im)}$ (= 29.2 Hz) for the imGCs is faster than $f_p^{(m)}$ (= 11.4 Hz) for the mGCs, as can be well seen in Fig.~\ref{fig:SSR1}(a1), mainly because for $x=1$ firing of the imGCs is much more active than that of the mGCs (resulting from high excitability of the imGCs).

However, as $x$ is decreased from 1 to 0, $f_p^{(im)}$ decreases to 0 rapidly due to rapid decrease in firing activity of the imGCs (resulting from their low excitatory innervation). On the other hand, $f_p^{(m)}$ increases to 18.6 Hz because of increase in firing activity of the mGCs (resulting from decrease in the feedback inhibition into the mGCs). We note that, $f_p^{(im)}$ and $f_p^{(m)}$ cross at $ x^* \sim 0.4$; for $x > x^*$  $f_p^{(im)} > f_p^{(m)},$ while for $x < x^*$  $f_p^{(m)} > f_p^{(im)}.$ In the case of whole GCs, the population frequency $f_p^{(w)}$ follows tendency of the larger one between $f_p^{(m)}$ and $f_p^{(im)}$.
Thus, $f_p^{(w)}$ forms a well-shaped curve (i.e., $f_p^{(w)}$ decreases from 29.2 Hz to $\sim 18.1$ Hz for $ 1 \geq x > x^*$, while it increases for $x^* > x \geq 0$, and converges to a limit value (= 18.6 Hz) for $x=0$ which is a little larger than the dashed horizontal line ($f_p= 18.2$ Hz) in the homogeneous population  of only the mGCs (without imGCs), which may be understood as follows. In the limiting case of $x=0,$ the imGCs become completely inactive. Hence, the feedback inhibition (from the BCs and the HIPP cells) to the mGCs becomes reduced in comparison to the homogeneous case consisting of only mGCs, which results in increased firing activity of the mGCs in the heterogeneous population of mGCs and imGCs.

The amplitude of the IPSR $R^{(X)}(t)$ may represent synchronization degree of the SSR. Thus, we characterize the synchronization degree of the SSRs of the mGCs ($X=m$), the imGCs ($X=im$), and the whole GCs ($X=w$) in terms of the amplitude measure ${\cal M}_a^{(X)}$ of Eq.~(\ref{eq:AM}), given by the time-averaged amplitude of $R^{(X)}(t).$ Figure \ref{fig:SSR2}(b) shows the plots of ${\cal M}_a^{(X)}$ versus $x$ for the mGCs (green solid circles), the imGCs (red open circles), and the whole GCs (blue crosses). For $x=1$ (i.e., high excitability of the imGCs), ${\cal M}_a^{(im)}$ (= 2.75) for the imGCs is larger than ${\cal M}_a^{(m)}$ (= 0.99) for the mGCs, as can be seen well in Fig.~\ref{fig:SSR1}(a1), because the imGCs fire more actively and coherently than mGCs.

As $x$ is decreased from 1 to 0 (i.e., low excitatory innervation to the imGCs) the effect of imGCs becomes decreased rapidly, which results in more active and coherent firing activity of the mGCs (due to decreased feedback to the mGCs from the BCs and the HIPP cells). Consequently, ${\cal M}_a^{(m)}$ increases to 3.79, while ${\cal M}_a^{(im)}$ decreases to 0. In the case of whole GCs, ${\cal M}_a^{(w)}$ increases from 2.37 to 3.79 by following tendency of the larger one between ${\cal M}_a^{(m)}$ and ${\cal M}_a^{(im)}$, as can be well seen in Fig.~\ref{fig:SSR2}(b). We note that the limit value (= 3.79) of both ${\cal M}_a^{(m)}$ and ${\cal M}_a^{(w)}$ is a little smaller than ${\cal M}_a$ (= 3.83) in the homogeneous population  of only the mGCs (without imGCs) [represented by
the dashed horizontal line in Fig.~\ref{fig:SSR2}(b)]. Hence, for all $x,$ ${\cal M}_a^{(X)}$ of the mGCs, the imGCs, and the whole GCs is less than that (= 3.83) in the homogeneous case consisting of only mGCs. Consequently, in the whole range of $x$, due to heterogeneity caused by the imGCs, population firing behaviors (characterized in terms of ${\cal M}_a^{(X)}$) of mGCs, imGCs, and whole GCs in their SSRs become deteriorated, in comparison to that in the presence of only mGCs (without imGcs).

Next, we consider the individual firing behaviors of the active mGCs, imGCs, and whole GCs which are well shown in their ISI histograms.
Figure \ref{fig:SSR2}(c) shows the plots of the population-averaged MFRs $\langle f_i^{(X)} \rangle$ of the
individual mGCs (green solid circles), imGCs (red open circles), and whole GCs (blue crosses); $\langle f_i^{(X)} \rangle$ corresponds to the reciprocal
of the population-averaged ISI (${\langle {\rm ISI} \rangle}^{(X)}$) (i.e., ${\langle f_i^{(X)} \rangle} =  {1/{\langle {\rm ISI} \rangle}}^{(X)}$)
in the population-averaged ISI histogram of the $X$-population.

For $x=1$ $\langle f_i^{(im)} \rangle$ (= 28.7 Hz) for the imGCs is much faster than $\langle f_i^{(m)} \rangle$ (= 1.12 Hz) for the mGCs, as can be well seen in Fig.~\ref{fig:SSR1}(b1). In this case, due to their high excitability, the imGCs exhibit active firing activity, while the mGCs show very intermittent spikings due to strong feedback inhibition (from the BCs and the HIPP cells). Thus, in the case of mGCs the population-averaged MFR $\langle f_i^{(m)} \rangle$ is much less than the population frequency $f_p^{(m)}~(= 11.4$ Hz) for the SSR, due to random spike skipping, which is in contrast to the case of full synchronization where the population-averaged MFR is the same as the population frequency. On the other hand, in the case of imGCs, their population-averaged MFR $\langle f_i^{(im)} \rangle$ is close to the population frequency $f_p^{(im)}~(= 29.2$ Hz) for the SSR, and hence the active imGCs show nearly fully synchronized rhythm (i.e., most of all active imGCs fire in each spiking stripe) in the case of $x=1$.

However, as $x$ is decreased from 1 to 0, firing activity of imGCs is decreased rapidly due to low excitatory innervation. Consequently,
$\langle f_i^{(im)} \rangle$ decreases so rapidly from 28.7 Hz to 0. Thus, for $x < 1$, active imGCs distinctly exhibit random spike skipping, leading to
SSR with $f_p^{(im)} > {\langle f_i^{(im)} \rangle}$. On the other hand, with decreasing $x$ from 1, $\langle f_i^{(m)} \rangle$ of the mGCs increases slowly from 1.12 to 2.30 Hz, because of decrease in feedback inhibition to the mGCs. When passing a threshold ($\sim 0.3$), $\langle f_i^{(m)} \rangle$ crosses the horizontal dashed line (= 2.12 Hz), representing the population-averaged MFR in the presence of only the mGCs (without the imGCs). It also crosses the decreasing curve of
$\langle f_i^{(im)} \rangle$ for $x \sim 0.2$, and then converges to the limit value (= 2.30 Hz).

We also consider the case of whole GCs. For $x=1$, their population-averaged MFR $\langle f_i^{(w)} \rangle$ (= 24.1 Hz) is high due to the effect of the imGCs. However, with decreasing $x$ from 1 the effect of the imGCs is decreased, and hence $\langle f_i^{(w)} \rangle$ is decreased until $ x \sim 0.1$. Then, for $x < 0.1$ $\langle f_i^{(w)} \rangle$ begins to increase slowly and approach $\langle f_i^{(m)} \rangle$. Thus, in the limiting case of $x=0,$
$\langle f_i^{(w)} \rangle = \langle f_i^{(w)} \rangle =$ 2.30 Hz which is larger than the population-averaged MFR (= 2.12 Hz) in the case of homogeneous population
of only mGCs (without imGCs). For $x=0$ the imGCs become completely inactive. Hence, the feedback inhibition to the mGCs becomes decreased in comparison with the homogeneous case composed of only mGCs, which leads to increase in firing activity of the mGCs in the heterogeneous population of mGCs and imGCs.

Next, we characterize the degree of random spike skipping seen in the multi-peaked ISI histogram in the case of $X=m,~im,$ or $w$ in terms of the random phase-locking degree ${\cal L}_d^{(X)}$ of Eq.~(\ref{eq:LD}) (denoting how well intermittent spikes make phase-locking to the IPSR $R^{(X)}(t)$ at random multiples of its global period $T_G^{(X)}$). The sharper the random-spike-skipping peaks in the ISI histogram are, the larger ${\cal L}_d^{(X)} $ becomes.

Figure \ref{fig:SSR2}(d) shows the plots of ${\cal L}_d^{(X)}$ versus $x$ for the mGCs ($X=m:$ green solid circles), the imGCs ($X=im:$ red open circles), and the whole GCs ($X=w:$ blue crosses). In the case of mGCs, multi-peaked ISI histograms appear due to random spike-skippings. As $x$ is decreased from 1,
their ISI histograms become clearer because multiple peaks become sharper and their heights become increased. Thus, with decreasing $x$ from 1 to 0,
${\cal L}_d^{(m)}$ is found to increase from 0.587 to 0.898. On the other hand, the ISI histograms for the imGCs have single peaks. As $x$ is decreased from 1, their single-peaked ISI histograms become broader, which leads to more random spike skippings. But, at first, in the $n$th global cycle where the single peak exists [e.g., $n=1$ (2) for $x=1.0$ (0.7)], the random phase locking degree ${\cal L}_d^{(im)}(n)$ increases a little until $x$ is decreased to $x^* (\sim 0.7)$. Thus,
as $x$ is decreased from 1 to $x^*,$ the overall ${\cal L}_d^{(im)}$ increases a little from 0.408 to 0.497. Then, for $x < x^*$, ${\cal L}_d^{(im)}$ decreases rapidly to 0, in contrast to the case of mGCs.

In the case of whole GCs, both the peak (associated with the imGCs) and the multiple peaks (related to the mGCs) coexist in the histogram. Fractions of the peaks vary depending on $x$. For $x=1$, dominant major peak, related to the imGCs, appears, while fractions of multiple peaks, associated with the mGCs, are very small. Thus, the value of ${\cal L}_d^{(w)}$ is close to that of the imGCs. But, with decreasing $x$ from 1, the effect of the imGCs becomes weakened, and fractions of the multiple peaks, associated with the mGCs, become increased. Thus, as $x$ is decreased from 1, ${\cal L}_d^{(w)}$ is found to increase from 0.457 and converge
to ${\cal L}_d^{(m)}$ of the mGCs. Thus, in the limiting case of $x=0,$ ${\cal L}_d^{(w)} = {\cal L}_d^{(m)} = 0.898$ which is smaller than
${\cal L}_d$ (= 0.92) in the homogeneous population  of only the mGCs (without imGCs) [represented by the dashed curve in Fig.~\ref{fig:SSR2}(d)].
Hence, for all $x$ ${\cal L}_d^{(X)}$ of the mGCs, the imGCs, and the whole GCs is smaller than that (= 0.92) in the homogeneous case composed of only mGCs.
As a result, in the whole range of $x$, because of heterogeneity caused by the imGCs, individual firing behaviors (characterized in terms of
${\cal L}_d^{(X)}$) of mGCs, imGCs, and whole GCs in their SSRs become deteriorated, in comparison with the homogeneous case consisting of only mGCs (without imGcs).

\begin{figure}
\includegraphics[width=\columnwidth]{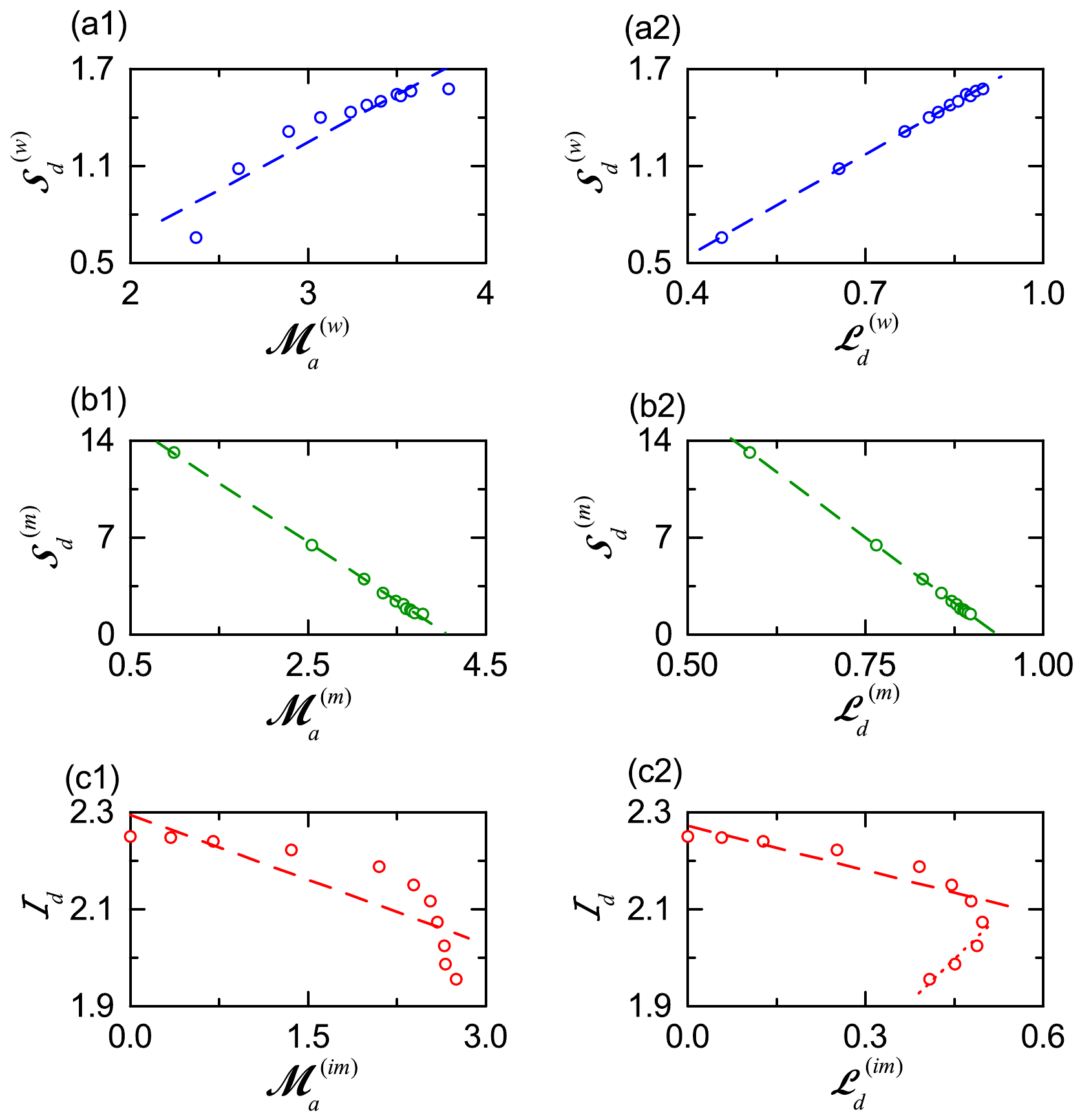}
\caption{Quantitative relationship between SSRs and pattern separation and integration in each case of mGCs, imGCs, and whole GCs.
In the case of whole GCs, plots of (a1) the pattern separation degree ${\cal S}_d^{(w)}$ versus the amplitude measure ${\cal M}_a^{(w)}$ and (a2) ${\cal S}_d^{(w)}$ versus the random phase-locking degree ${\cal L}_d^{(w)}.$ In the case of mGCs, plots of (b1) the pattern separation degree ${\cal S}_d^{(m)}$ versus ${\cal M}_a^{(m)}$ and (b2) ${\cal S}_d^{(m)}$ versus ${\cal L}_d^{(m)}.$ In the case of imGCs, plots of (c1) the pattern integration degree ${\cal I}_d$ versus ${\cal M}_a^{(im)}$ and (c2) ${\cal I}_d$ versus ${\cal L}_d^{(im)}.$ Fitted dashed lines are given in (a1)-(c1). In (c2), fitted dashed and dotted lines are obtained from 8 data points for $0.7 \leq x \leq 0$ and 4 data points for $1 \leq x \leq 0.7$, respectively. The mGCs, imGCs, and whole GCs are denoted by green solid circles, red open circles, and blue crosses, respectively.
}
\label{fig:QR}
\end{figure}

Finally, we investigate quantitative association between SSRs and pattern separation and integration.
In \cite{NG-PS}, by varying the synaptic connectivity fraction $x$, we studied pattern separation (transforming similar input patterns into less similar output patterns) of the mGCs ($X=m$) and the whole GCs ($X=w$) in terms of the pattern separation degree ${\cal S}_d^{(X)}$ [see Fig.~4(e) in \cite{NG-PS}]. The pattern separation degree ${\cal S}_d^{(X)}$ is given by the ratio of $D_p^{(out)}$ (pattern distance for the output pattern pair) to $D_p^{(in)}$ (pattern distance for the input pattern pair) [see Eq.~(17) in \cite{NG-PS}]. For ${\cal S}_d^{(X)} >1$ pattern separation occurs, because the output pattern pair of the mGCs is more dissimilar than the input pattern pair of the EC cells. On the other hand, for ${\cal S}_d^{(X)} <1$ no pattern separation occurs; instead, pattern convergence takes place. For the mGCs, with decreasing $x$ from 1, ${\cal S}_d^{(m)}$ was found to decrease from a high value (= 13.142) for $x=1$ to a limit value (= 1.495) for $x=0$, as shown in Fig.~4(e) in \cite{NG-PS}. Thus, in the whole range of $ 0 \leq x \leq 1$, the mGCs perform good pattern separation with ${\cal S}_d^{(m)} > 1$.

In contrast to the mGCs, the imGCs exhibit pattern integration (making association between patterns), characterized in terms of the integration degree
${\cal I}_d$. ${\cal I}_d$ is given by the average pattern correlation degree for the output pattern pair of the imGCs to the average pattern correlation degree for the input pattern pair of the EC cells [see Eq.~(18) in \cite{NG-PS}]. For $x = 1$ the pattern integration degree ${\cal I}_d$ of the imGCs is high (1.9559). With decreasing $x$ from 1 to 0, ${\cal I}_d$ was found to increase from 1.9559 to 2.2502, as shown in Fig.~4(f) in \cite{NG-PS}. Thus, in the whole range of $ 0 \leq x \leq 1$, the imGCs are good pattern integrators with ${\cal I}_d >1.$

Then, we consider the heterogeneous population of whole GCs (composed of mGCs and imGCs). In this heterogeneous whole population, ${\cal S}_d^{(w)}$ for $x=1$ is 0.659. Since ${\cal S}_d^{(w)} <1$ for $x=1$, no pattern separation occurs, due to strong correlations between the imGCs. As $x$ is decreased from 1, the effect of the imGCs becomes reduced. Thus, when decreasing through a threshold $x^*$ (= 0.92), pattern separation (with ${\cal S}_d^{(w)} >1$) begins, and then the overall pattern separation degree ${\cal S}_d^{(w)}$ increases and converges to a limit value (= 1.577) for $x=0$ [see Fig.~4(e) in \cite{NG-PS}].

Figures \ref{fig:QR}(a1)-\ref{fig:QR}(a2) and \ref{fig:QR}(b1)-\ref{fig:QR}(b2) show the plots of the pattern separation degree ${\cal S}_d^{(X)}$ versus ${\cal M}_a^{(X)}$ and ${\cal S}_d^{(X)}$ versus ${\cal L}_d^{(X)}$ for $X=w$ (whole GCs) and $m$ (mGCs), respectively.
In the whole population of all the GCs, ${\cal S}_d^{(w)}$ is found to be positively correlated with ${\cal M}_a^{(w)}$ and ${\cal L}_d^{(w)}$ of their SSR
with the Pearson's correlation coefficients $r=0.9289$ and $0.9997$, respectively \cite{Pearson}. Thus, in the whole population of all the GCs, the larger ${\cal M}_a^{(w)}$ and ${\cal L}_d^{(w)}$ in the SSR are, the better pattern separation efficacy becomes. On the other hand, ${\cal S}_d^{(m)}$ for the mGCs are found to be negatively correlated with ${\cal M}_a^{(X)}$ and ${\cal L}_d^{(X)}$ of their SSR with the Pearson's correlation coefficients $r=- 0.9994$ and $- 0.9999$, respectively. Thus, in the population of the mGCs, the better population and individual firing behaviors in their SSR are, the worse their pattern separation   becomes.

In the case of $X=im$ (imGCs), plots of the pattern integration degree ${\cal I}_d$ versus ${\cal M}_a^{(im)}$ and ${\cal I}_d$ versus ${\cal L}_d^{(im)}$ are shown in Figs.~\ref{fig:QR}(c1)-\ref{fig:QR}(c2). As in the case of mGCs, ${\cal I}_d$ for the imGCs is also negatively correlated with ${\cal M}_a^{(im)}$ with the Pearson's correlation coefficients $r=- 0.8483$. Next we consider quantitative relationship between ${\cal I}_d$ and ${\cal L}_d^{(im)}$.
As shown in Fig.~\ref{fig:SSR2}(d), for $1.0 \geq x \geq 0.7$, ${\cal L}_d^{(im)}$ makes a small increase, and then for $x \leq 0.7$ it rapidly decreases to 0.
Thus, ${\cal I}_d$ is also negatively correlated with ${\cal L}_d^{(im)}$ in the range of $0.7 \geq x \geq 0$ with the Pearson's correlation coefficients
$r= - 0.9159$, as in the case of  ${\cal M}_a^{(im)}$, while in the initial small range of $1.0 \geq x \geq 0.7,$ ${\cal I}_d$ is positively correlated with ${\cal L}_d^{(im)}$ with the Pearson's correlation coefficients $r=~0.9365$. Thus, in the population of the imGCs, for $0.7 \geq x \geq 0$ the better population and individual firing behaviors in their SSR are, the worse their integration becomes.

\section{Summary and Discussion}
\label{sec:SUM}
We studied the effect of adult neurogenesis on the SSRs of the GCs (mGCs, imGCs, and whole GCs) in our DG spiking neural network. In comparison to the mGCs, the imGCs show two competing distinct properties of high excitability (causing high activation) and low excitatory innervation (reducing activation degree). Thus, the effect of low excitatory innervation counteracts the effect of high excitability.
The connection probability $p_c$ from the EC cells and the MCs to the imGCs is $20~x~\%$ [$x$ (synaptic connectivity fraction); $0 \leq x \leq 1$], in contrast to the case of mGCs with $p_c = 20~\%$. With decreasing $x$ from 1 to 0, population and individual firing behaviors of the mGCs, the imGCs, and the whole GCs in their SSRs  were characterized in terms of the amplitude measure ${\cal M}_a^{(X)}$ ($X=m,~im,$ and $w$) and the random phase-locking degree ${\cal L}_d^{(X)}$, respectively.

As shown in Fig.~\ref{fig:SSR2}, as $x$ is decreased from 1, the amplitude measure ${\cal M}_a^{(X)}$ and the random phase-locking degree ${\cal L}_d^{(X)}$ were found to increase in the case of mGCs ($X=m$) and whole GCs ($X=w$). With decreasing $x$ from 1, the effect of the imGCs became weaker, which resulted in increase in ${\cal M}_a^{(X)}$ and ${\cal L}_d^{(X)}$ for the mGCs and the whole GCs. In contrast, as $x$ decreases from 1, ${\cal M}_a^{(im)}$ of the imGCs was found to monotonically decrease, and their ${\cal L}_d^{(im)}$ was found to first slowly a little increase and then rapidly decrease to zero (i.e., for $x< 0.7$,
monotonic decrease to zero occurs). In this way, the changing tendency for the imGCs was in contrast to those of the mGCs and the whole GCs.
We also note that in the heterogeneous population (consisting of the mGCs and the imGCs), ${\cal M}_a^{(X)}$ and ${\cal L}_d^{(X)}$ ($X=m,$ $im,$ and $w$) were less than those in the homogeneous population of only mGCs without imGCs. Due to heterogeneity caused by the imGCs, the population and individual firing behaviors of the
GCs in the SSRs became deteriorated, in comparison with that in the presence of only mGCs (without imGCs).

Previously, in the whole range of $x$ (i.e., for $1 \geq x \geq 0$), the mGCs and the imGCs were found to exhibit pattern separation and pattern integration, respectively \cite{NG-PS}. As $x$ is decreased from 1, the pattern separation degree ${\cal S}_d^{(m)}$ of the mGCs was found to decrease, as shown in Fig.~4(e) in \cite{NG-PS}, because their activation degree increased. In contrast to the mGCs, the imGCs was found to show pattern integration, and its degree ${\cal I}_d$ was found to increase as $x$ is decreased from 1, due to increase in correlation between the imGCs [see Fig.~4(f) in \cite{NG-PS}. Due to presence of imGCs (good pattern integrators), the pattern separation efficacy in the whole heterogeneous population (composed of mGCs and imGCs) was also found to become deteriorated, as in the case of the SSR.
 
Quantitative association between SSRs and pattern separation and integration was shown in Fig.~\ref{fig:QR}.
In the whole population of all the GCs, ${\cal S}_d^{(w)}$ was found to be positively correlated with ${\cal M}_a^{(w)}$ and ${\cal L}_d^{(w)}$ of their SSR.
Hence, in the whole population, the larger ${\cal M}_a^{(w)}$ and ${\cal L}_d^{(w)}$ in the SSR are, the better the pattern separation efficacy becomes, as in the homogeneous population of only mGCs (without imGCs) \cite{PS}.
On the other hand,  ${\cal S}_d^{(m)}$ for the mGCs was found to be negatively correlated with ${\cal M}_a^{(m)}$ and ${\cal L}_d^{(m)}$ of their SSR.
Thus, in the population of the mGCs, the better population and individual firing behaviors in their SSR are, the worse their pattern separation efficacy becomes.
Also, in the case of imGCs, for $0.7 \geq x \geq 0,$ ${\cal I}_d$ was found to be negatively correlated to ${\cal M}_a^{(im)}$ and ${\cal L}_d^{(im)}$ of their SSR.
Hence, in the population of imGCs, for $0.7 \geq x \geq 0,$ the better population and individual firing behaviors in their SSR are, the worse their pattern integration efficacy becomes.

Finally, we discuss limitations of our present work and future works.
In the present work, although correlations between the pattern separation and integration degrees and the population synchronization and random phase-locking degrees in the SSRs were found, this kind of correlations do not imply causal relationship. Hence, in future work, it would be interesting to intensively investigate their dynamical causation.

We also note that the pyramidal cells in the CA3 provide inhibitory backprojections to the mGCs via polysynaptic connections, mediated by the BCs and the HIPP cells \citep{Myers2,Myers3,Scharfman}. These inhibitory backprojections may decrease the activation degree of the mGCs, resulting in improvement of pattern separation in the population of the mGCs. Hence, as a future work, it would be meaningful to study the effects of the backprojections on pattern separation and SSR in the combined DG-CA3 network.

In the whole heterogeneous population of all the GCs (mGCs and imGCs), both the pattern separation efficacy and
the regularity of population and individual firing activities in the SSR were found to get  deteriorated, due to presence of the imGCs (pattern integrators).
But, we note that the pattern separation may not always be a strict requirement for accurate neural encoding.
In the homogeneous population of only the mGCs (without the imGCs), memory storage capacity could be increased with pattern separation efficacy \cite{Myers1}.
On the other hand, in a heterogeneous population of mGCs (pattern separators) and imGCs (pattern integrators), the memory storage capacity might be optimally maximized via mixed encoding via pattern separation on similar input patterns and pattern integration on very dissimilar input patterns \cite{NG9,Hetero,NG-PS}.
Thus, through mixed encoding, memory resolution (corresponding to the extent of information incorporated into memories) could be increased, which would result in reduction in memory interference, although regularity of population and individual firing activities in the SSR becomes deteriorated.
This speculation on increase in memory resolution via mixed encoding (through cooperation of pattern separation and pattern integration) must be examined in future works.

\section*{Acknowledgments}
This research was supported by the Basic Science Research Program through the National Research Foundation of Korea (NRF) funded by the Ministry of Education (Grant No. 20162007688).

\appendix
\section{Leaky Integrate-and-Fire Models for Single Spiking Neurons}
\label{app:A}
As elements of our DG spiking neural network, we choose LIF spiking neuron models with additional AHP currents (determining the refractory period).
The following equations govern evolution of dynamical states of individual cells in the $X$ population:
\begin{eqnarray}
C_{X} \frac{dv_{i}^{(X)}(t)}{dt} &=& -I_{L,i}^{(X)}(t) - I_{AHP,i}^{(X)}(t) + I_{ext}^{(X)} - I_{syn,i}^{(X)}(t), 
\nonumber \\
& & \;\;\; i=1, \cdots, N_{X}, 
\label{eq:GE}
\end{eqnarray}
where $N_X$ is the total number of cells in the $X$ population, $X=$ mGC, imGC, and BC in the granular layer and $X=$ MC and HIPP cell in the hilus.
In Eq.~(\ref{eq:GE}), $C_{X}$ (pF) represents the membrane capacitance of the cells in the $X$ population, and the dynamical state of the $i$th cell in the $X$ population at a time $t$ (msec) is characterized by its membrane potential $v_i^{(X)}(t)$ (mV). We note that the time-evolution of $v_i^{(X)}(t)$ is governed by 4 types of currents (pA) into the $i$th cell in the $X$ population; the leakage current $I_{L,i}^{(X)}(t)$, the AHP current $I_{AHP,i}^{(X)}(t)$, the external constant current $I_{ext}^{(X)}$ (independent of $i$), and the synaptic current $I_{syn,i}^{(X)}(t)$.
Here, we consider a subthreshold case of $I_{ext}^{(X)}=0$  for all $X$ \cite{Chavlis}.

In Eq.~(\ref{eq:GE}), the 1st type of leakage current $I_{L,i}^{(X)}(t)$ for the $i$th neuron in the $X$ population is given by:
\begin{equation}
I_{L,i}^{(X)}(t) = g_{L}^{(X)} (v_{i}^{(X)}(t) - V_{L}^{(X)}),
\label{eq:Leakage}
\end{equation}
where $g_L^{(X)}$ and $V_L^{(X)}$ are conductance (nS) and reversal potential for the leakage current, respectively.
When its membrane potential $v_i^{(X)}$ reaches a threshold $v_{th}^{(X)}$ at a time $t_{f,i}^{(X)}$, the $i$th neuron in the $X$ population
fires a spike. After spiking (i.e., $t \geq t_{f,i}^{(X)}$), the 2nd type of AHP current $I_{AHP,i}^{(X)}(t)$ follows:
\begin{equation}
I_{AHP,i}^{(X)}(t) = g_{AHP}^{(X)}(t) ~(v_{i}^{(X)}(t) - V_{AHP}^{(X)})~~~{\rm ~for~} \; t \ge t_{f,i}^{(X)}.
\label{eq:AHP1}
\end{equation}
Here, $V_{AHP}^{(X)}$ is the reversal potential for the AHP current, and the conductance $g_{AHP}^{(X)}(t)$ is given by an exponential-decay
function:
\begin{equation}
g_{AHP}^{(X)}(t) = \bar{g}_{AHP}^{(X)}~  e^{-(t-t_{f,i}^{(X)})/\tau_{AHP}^{(X)}} ,
\label{eq:AHP2}
\end{equation}
where $\bar{g}_{AHP}^{(X)}$ and $\tau_{AHP}^{(X)}$ are the maximum conductance and the decay time constant for the AHP current.
With increasing $\tau_{AHP}^{(X)}$, the refractory period becomes longer.

\section{Three Types of Synaptic Currents}
\label{app:B}

We consider the 4th type of synaptic current $I_{syn,i}^{(X)}(t)$ into the $i$th neuron in the $X$ population in Eq.~(\ref{eq:GE}).
The synaptic current $I_{syn,i}^{(X)}(t)$ consists of the following 3 kinds of synaptic currents:
\begin{equation}
I_{syn,i}^{(X)}(t) = I_{{\rm AMPA},i}^{(X,Y)}(t) + I_{{\rm NMDA},i}^{(X,Y)}(t) + I_{{\rm GABA},i}^{(X,Z)}(t).
\label{eq:ISyn1}
\end{equation}
Here, $I_{{\rm AMPA},i}^{(X,Y)}(t)$ and $I_{{\rm NMDA},i}^{(X,Y)}(t)$ are the excitatory AMPA ($\alpha$-amino-3-hydroxy-5-methyl-4-isoxazolepropionic acid) receptor-mediated and NMDA ($N$-methyl-$D$-aspartate) receptor-mediated currents from the presynaptic source $Y$ population to the postsynaptic $i$th neuron in the target $X$ population. On the other hand, $I_{{\rm GABA},i}^{(X,Z)}(t)$ is the inhibitory $\rm GABA_A$ ($\gamma$-aminobutyric acid type A) receptor-mediated current
from the presynaptic source $Z$ population to the postsynaptic $i$th neuron in the target $X$ population.

Like in the case of the AHP current, the $R$ (= AMPA, NMDA, or GABA) receptor-mediated synaptic current $I_{R,i}^{(T,S)}(t)$ from the presynaptic source $S$ population to the $i$th postsynaptic neuron in the target $T$ population is given by:
\begin{equation}
I_{R,i}^{(T,S)}(t) = g_{R,i}^{(T,S)}(t)~(v_{i}^{(T)}(t) - V_{R}^{(S)}),
\label{eq:ISyn2}
\end{equation}
where $g_{(R,i)}^{(T,S)}(t)$ and $V_R^{(S)}$ are synaptic conductance and synaptic reversal potential
(determined by the type of the presynaptic source $S$ population), respectively.

In the case of the $R$ (=AMPA and GABA)-mediated synaptic currents, we get the synaptic conductance $g_{R,i}^{(T,S)}(t)$ from:
\begin{equation}
g_{R,i}^{(T,S)}(t) = K_{R}^{(T,S)} \sum_{j=1}^{N_S} w_{ij}^{(T,S)} ~ s_{j}^{(T,S)}(t),
\label{eq:ISyn3}
\end{equation}
where $K_{R}^{(T,S)}$ is the synaptic strength per synapse for the $R$-mediated synaptic current
from the $j$th presynaptic neuron in the source $S$ population to the $i$th postsynaptic cell in the target $T$ population.

On the other hand, in the NMDA-receptor case, some of the postsynaptic NMDA channels are blocked by the positive magnesium ion ${\rm Mg}^{2+}$
\cite{NMDA}. Hence, the conductance in the case of NMDA receptor is given by \cite{Chavlis}:
\begin{equation}
g_{R,i}^{(T,S)}(t) = {\widetilde K}_R^{(T,S)}~f(v^{(T)}(t))~\sum_{j=1}^{N_S} w_{ij}^{(T,S)} ~ s_{j}^{(T,S)}(t).
\label{eq:NMDA}
\end{equation}
Here, ${\widetilde K}_R^{(T,S)}$ is the synaptic strength per synapse, and fraction of NMDA channels that are not blocked by the ${\rm Mg}^{2+}$ ion is given by a sigmoidal function $f(v^{(T)}(t))$:
\begin{equation}
f(v^{(T)}(t)) = \frac{1}{1+\eta\cdot [{\rm Mg}^{2+}]_o \cdot \exp(-\gamma \cdot v^{(T)}(t))}.
\end{equation}
Here, $v^{(T)}(t)$ is the membrane potential of the target cell, $[{\rm Mg}^{2+} ]_o$ is the outer ${\rm Mg}^{2+}$ concentration, $\eta$ denotes the sensitivity of ${\rm Mg}^{2+}$ unblock, $\gamma$ represents the steepness of ${\rm Mg}^{2+}$ unblock, and the values of parameters change depending on the target cell \cite{Chavlis}.
For simplicity, some approximation to replace $f(v^{(T)}(t))$ with $\langle f(v^{(T)}(t))\rangle$ [i.e., time-averaged value of $f(v^{(T)}(t))$ in the range of $v^{(T)}(t)$ of the target cell] has been done in \cite{SSR}. Then, an effective synaptic strength $K_{\rm NMDA}^{(T,S)} (={\widetilde K}_{\rm NMDA}^{(T,S)} \langle f(v^{(T)}(t))\rangle$) was introduced by absorbing $\langle f(v^{(T)}(t))\rangle$ into $K_{\rm NMDA}^{(T,S)}$. Thus, with the scaled-down effective synaptic strength $K_{\rm NMDA}^{(T,S)}$ (containing the blockage effect of the ${\rm Mg}^{2+}$ ion), the conductance $g$ for the NMDA receptor may also be well approximated in the same form of conductance as the other AMPA and GABA receptors in Eq.~(\ref{eq:ISyn3}). Thus, we get all the effective synaptic strengths $K_{\rm NMDA}^{(T,S)}$ from the synaptic strengths $\widetilde{K}_{\rm NMDA}^{(T,S)}$ in \cite{Chavlis} by considering the average blockage effect of the ${\rm Mg}^{2+}$ ion.
Consequently, we can use the same form of synaptic conductance of Eq.~(\ref{eq:ISyn3}) in all the cases of $R=$ AMPA, NMDA, and GABA.

The interpopulation synaptic connection from the source $S$ population (with $N_s$ neurons) to the target $T$ population is given by the connection weight matrix
$W^{(T,S)}$ ($=\{ w_{ij}^{(T,S)} \}$) where $w_{ij}^{(T,S)}=1$ if the $j$th neuron in the source $S$ population is pre-synaptic to the $i$th neuron
in the target $T$ population; otherwise $w_{ij}^{(T,S)}=0$.

The postsynaptic ion channels are opened through binding of neurotransmitters (emitted from the source $S$ population) to receptors in the target
$T$ population. Fraction of open ion channels at time $t$ is represented by $s^{(T,S)}(t)$. The time course of $s_j^{(T,S)}(t)$ of the $j$th cell
in the source $S$ population is given by a sum of double exponential functions $E_{R}^{(T,S)} (t - t_{f}^{(j)}-\tau_{R,l}^{(T,S)})$:
\begin{equation}
s_{j}^{(T,S)}(t) = \sum_{f=1}^{F_{j}^{(s)}} E_{R}^{(T,S)} (t - t_{f}^{(j)}-\tau_{R,l}^{(T,S)}).
\label{eq:ISyn4}
\end{equation}
Here, $t_f^{(j)}$ and $F_j^{(s)}$ are the $f$th spike time and the total number of spikes of the $j$th cell in the source $S$ population, respectively, and
$\tau_{R,l}^{(T,S)}$ is the synaptic latency time constant for $R$-mediated synaptic current.
The exponential-decay function $E_{R}^{(T,S)} (t)$ (corresponding to contribution of a presynaptic spike occurring at $t=0$ in the absence of synaptic latency)
is given by:
\begin{equation}
E_{R}^{(T,S)}(t) = \frac{1}{\tau_{R,d}^{(T,S)}-\tau_{R,r}^{(T,S)}} \left( e^{-t/\tau_{R,d}^{(T,S)}} - e^{-t/\tau_{R,r}^{(T,S)}} \right) \cdot \Theta(t). \label{eq:ISyn5}
\end{equation}
Here, $\Theta(t)$ is the Heaviside step function: $\Theta(t)=1$ for $t \geq 0$ and 0 for $t <0$, and $\tau_{R,r}^{(T,S)}$ and $\tau_{R,d}^{(T,S)}$ are synaptic rising and decay time constants of the $R$-mediated synaptic current, respectively.

In comparison with our prior DG networks \cite{WTA,SSR,PS}, we include more synaptic connections with a high degree of anatomical and physiological realism \cite{BN1,BN2}, and incorporate the imGCs. Thus, a new feedforward inhibition, mediated by the BCs, is provided to the mGCs, and
there appear two feedback loops of mGC-BC and mGC-HIPP, (projecting feedback inhibition to the mGCs), the activities of which
are controlled by the two control loops of MC-BC and MC-HIPP (MCs: controllers).

\end{document}